\documentclass{aastex631}
\usepackage{amsmath}
\usepackage{bm}
\usepackage{lmodern}
\shorttitle{AASTeX v6.3.1 Sample article}
\shortauthors{Wang et al.}
\graphicspath{{./}{figures/}}

\begin{document}
\title{Numerical simulation of oscillatory magnetic reconnection modulated by solar convective motions}

\author{Yifu Wang}
\affiliation{Yunnan Observatories, \\
Chinese Academy of Sciences, \\
Kunming, \\
Yunnan 650216, People’s Republic of China;}
\affiliation{University of Chinese Academy of Sciences, \\
Beijing 100049, \\
People’s Republic of China;}

\author[0000-0001-6366-7724]{Lei Ni}
\affiliation{Yunnan Observatories, \\
Chinese Academy of Sciences, \\
Kunming, \\
Yunnan 650216, People’s Republic of China;}
\affiliation{University of Chinese Academy of Sciences, \\
Beijing 100049, \\
People’s Republic of China;}
\affiliation{Center for Astronomical Mega-Science, \\
Chinese Academy of Sciences, \\
20A Datun Road, \\
Chaoyang District, Beijing 100012,\\
People’s Republic of China;}
\email{leini@ynao.ac.cn}

\author{Guanchong Cheng}
\affiliation{Yunnan Observatories, \\
Chinese Academy of Sciences, \\
Kunming, \\
Yunnan 650216, People’s Republic of China;}
\affiliation{University of Chinese Academy of Sciences, \\
Beijing 100049, \\
People’s Republic of China;}
\email{chengguanchong@ynao.ac.cn}

\author[0000-0001-9828-1549]{Jialiang Hu}
\affiliation{Yunnan Observatories, \\
Chinese Academy of Sciences, \\
Kunming, \\
Yunnan 650216, People’s Republic of China;}
\affiliation{University of Chinese Academy of Sciences, \\
Beijing 100049, \\
People’s Republic of China;}

\author{Yuhao Chen}
\affiliation{Yunnan Observatories, \\
Chinese Academy of Sciences, \\
Kunming, \\
Yunnan 650216, People’s Republic of China;}
\affiliation{University of Chinese Academy of Sciences, \\
Beijing 100049, \\
People’s Republic of China;}
\affiliation{Harvard-Smithsonian Center for Astrophysics, 60 Garden Street, Cambridge, MA 02138, USA;}

\author{Abdullah Zafar}
\affiliation{Yunnan Observatories, \\
Chinese Academy of Sciences, \\
Kunming, \\
Yunnan 650216, People’s Republic of China;}



\begin{abstract}
\   Oscillatory magnetic reconnection is a periodic magnetic reconnection process, during which the current sheet's orientation and the magnetic connections change periodically. This periodic variation is generally considered to originate from the magnetic reconnection itself rather than from external driving processes. We conduct 2.5-dimensional radiative magnetohydrodynamic simulations to investigate the emergence of a magnetic flux tube from the convection zone into the lower corona, where the emerging magnetic fields reconnect with background ones. During the reconnection process within 5771 s, the current sheet's orientation has been reversed 41 times, corresponding to 40 oscillation periods. Notably, the longest period is 30 minutes, which is consistent with the previous observational results. We find that the main factor leading to the reversal of the current sheet's orientation is the quasi-periodic external force provided by the emergence of plasma and magnetic fields from the convection zone. We also find the shifting of the upward outflows from the reconnection region along the horizontal direction due to the alternating changes of the reconnection inflow and outflow regions. In addition to the quasi-periodic change of the current sheet orientation, the reconnection rate at the main X-point also oscillates with a period between 100-400 s, which corresponds to the period of p-mode oscillations.

\end{abstract}




\section{Introduction}

\   Magnetic reconnection is a fundamental physical process in the solar atmosphere, responsible for the release of magnetic energy and dissipation of currents located between oppositely directed magnetic field lines.  \citet{1991ApJ...371L..41C} firstly identified the possibility of periodic variations in current sheet configurations during magnetic reconnection. In their 2D simulations, an initial vertical current sheet transformed into a horizontal one during a reconnection process, and then its orientation became vertical again. The oscillatory transformation of the current sheet orientation at the magnetic null point is accompanied by periodic changes in magnetic connectivity. They named this phenomenon oscillatory magnetic reconnection.

\   Due to the limitations of the observation precision and resolution of instruments, the direct observations about the reversal of the current sheet orientation remain sparse. For the first time, \citet{Hong2019} found that a horizontal current sheet transformed into a vertical one by analyzing data from the Solar Dynamics Observatory (SDO) \citep{Pesnell2012}. The durations of the two current sheets were 23 minutes and 7 minutes, respectively, corresponding to an oscillation period of 30 minutes. The reversal of reconnection outflow and inflow regions has also been confirmed by the New Vacuum Solar Telescope (NVST) \citep{Xue2019, Sun2023}. In the work by \citet{Xue2019}, they observed four current sheets appearing one after another almost at the same location, and the orientation of the first and third current sheets is almost perpendicular to the second and fourth current sheets. The measured periods were 206 minutes, 433 minutes, and 615 minutes \citep{Xue2019}, respectively.

\   Plenty of observational phenomena imply that the oscillatory reconnection might frequently occur in the solar atmosphere, though these observations did not directly show the reversal of the current sheet orientation. For example, in the extreme ultraviolet (EUV) passbands, the recurring jets originated from two different sites close to each other in NOAA active region 12035 have been observed showing parallel trajectories \citep{Joshi2017}. \citet{Mulay2017} also observed multiple intermittent parallel shifting jets with both cool emissions and hot coronal emissions. Furthermore, the intermittent jets along different directions have been observed in the H-alpha bands in the region with mixed-polarity magnetic fields \citep{Cai2024}. These different intermittent jets are presumed to be the consequences of oscillatory reconnection processes. However, we should also note that there is another type of oscillatory reconnection: the peak reconnection rates and current density can periodically vary with time, but the reversal of the current sheet orientation is not observed. The solar activities such as the periodic Type III radio bursts \citep{Cattell2021}, the quasi-periodic oscillations in emission intensities of small brightenings and solar flares \citep{Innes1997, Peter2014, Tian2016, Young2018, Ning2014, Li2020, Hayes2020, Zimovets2021}, the periodic radiation intensity of the current sheet \citep{Kumar2024}, and the periodic perturbations of coronal magnetic loops \citep{Zhang2020, Ning2022} might be related to such type of oscillatory reconnections.

\   Many numerical studies have focused on studying the driving mechanisms of the oscillatory reconnection. In the 2D magnetohydrodynamic (MHD) simulations by \citet{McLaughlin2009}, an artificially imposed velocity pulse near the magnetic X-point evolved into a shock wave, then it deformed the X-point structure into a cusp-like configuration, which subsequently collapsed into a horizontal current sheet. The hot jets ejected out from the horizontal current sheet then accumulated to form a back-pressure in the outflow region, which squashes the horizontal current sheet to transform into a vertically oriented one. Such oscillatory magnetic reconnection driven by hot reconnection outflows has also been observed in 3D numerical simulations with initial magnetic null points \citep{Thurgood2017}. The external oscillatory is not included to trigger the oscillatory process in these simulations (e.g., \citeauthor{McLaughlin2009} \citeyear{McLaughlin2009}, \citeauthor{Thurgood2017} \citeyear{Thurgood2017}). However, the solar atmosphere is normally very dynamic; the accumulated hot reconnection outflow is probably not the only mechanism to drive the oscillatory magnetic reconnection in the solar atmosphere.

\   The magnetic flux emergence and convective motions below and around the solar surface might significantly affect the reconnection process and even trigger the solar eruptions directly \citep{Cheung2014, Chen2000, Lin2001, shibata2007, Ni2021, Cheng2021, chenyh2022, chen2023}. In the previous 2.5D simulation (e.g., \citeauthor{Murray2009} \citeyear{Murray2009}, \citeauthor{McLaughlin2012a} \citeyear{McLaughlin2012a}), a magnetic flux rope was inserted below the solar surface, and an initial density perturbation was included to trigger the emergence of this flux rope, then the oscillatory magnetic reconnection happens between the emerged flux rope and the background magnetic fields. The continuous reconnection outflows cause the accretion of hot and high-density plasmas in the outflow region, where the high plasma pressure results in the alteration of the force balance between the outflow and inflow regions and drives the reversal of the current sheet orientation \citep{Murray2009}. Therefore, the driving mechanism of oscillatory reconnection between the background magnetic fields and the emerged magnetic flux ropes in those previous simulations (e.g., \citeauthor{Murray2009} \citeyear{Murray2009}, \citeauthor{McLaughlin2012a} \citeyear{McLaughlin2012a}, \citeauthor{Stewart2022} \citeyear{Stewart2022}) is still similar to that in the simple initial X-point structure reconnection (e.g., \citeauthor{McLaughlin2009} \citeyear{McLaughlin2009}, \citeauthor{Thurgood2017} \citeyear{Thurgood2017}). It is important to note that previous simulations studying oscillatory magnetic reconnection have not considered the convection occurring beneath the photosphere and the resulting fluctuations in the solar atmosphere, which could be crucial in modulating the reconnection process in the lower solar atmosphere. (e.g, \citeauthor{Hong2022} \citeyear{Hong2022}).

\   The periods of current sheet orientation reversals (also called the periods of the oscillatory reconnection) can be affected by many factors. In the simulations of magnetic reconnection triggered by velocity pulses near the X-point \citep{McLaughlin2012b}, the results showed that a greater initial perturbation shortens the period of the oscillatory reconnection. However, in the coronal, the influences of the initial perturbation on this period have been weakened by the heat conduction, which can significantly speed up the dissipations of disturbances \citep{Karampelas2022a}. In addition, other important physical parameters, such as density, resistivity, and background magnetic field strength, can also affect the period of the oscillatory reconnection \citep{Thurgood2019, Karampelas2022b, Karampelas2023, Schiavo2024a, Schiavo2024b, Talbot2024}. In most previous simulations, the period of oscillatory reconnection is normally about several minutes (e.g., \citeauthor{McLaughlin2012a} \citeyear{McLaughlin2012a}, \citeauthor{McLaughlin2012b} \citeyear{McLaughlin2012b}, \citeauthor{Karampelas2022a} \citeyear{Karampelas2022a}, \citeauthor{Karampelas2022b} \citeyear{Karampelas2022b}, \citeauthor{Karampelas2023} \citeyear{Karampelas2023}), which is shorter than that observed in the solar atmosphere \citep{Hong2019, Xue2019}. In the work by \citet{Murray2009}, they mentioned that there is an oscillatory reconnection period lasting up to 30 minutes in the late reconnection phase while the current density has already been substantially reduced.

\   In this work, we have performed 2.5D Radiative MHD simulations to study magnetic reconnection between the emerged and background magnetic fields. The simulation domain extends from the upper convection zone to the low corona. The magnetic flux emergence and reconnection processes are modulated by convective and turbulent motions in the convection zone. We have analyzed the driving mechanism of multiple oscillatory reconnection processes, the oscillation periods, and the resulting parallel jets. In the following part of this paper, the numerical models are described in the second section. The third section presents the numerical results. The summaries and conclusions are given in the last section.

\section{Numerical Models}
\subsection{MHD model}
\   The developed NIRVANA code \citep{Ziegler2008, Ni2022, Cheng2024}  was used to perform the numerical simulations. The solved MHD equations are as follows:
\begin{equation}
\frac{\partial\rho}{\partial t}=-\nabla\cdot(\rho\boldsymbol{v}),
\end{equation}
\begin{equation}
\frac{\partial(\rho \boldsymbol{v})}{\partial t}=-\nabla\cdot[\rho \boldsymbol{vv}+(p+\frac{1}{2\mu_0}|\boldsymbol{B}|^{2})I-\frac{1}{\mu_0}\boldsymbol{BB}]+\rho \boldsymbol{g},
\end{equation}
\begin{equation}
\frac{\partial e}{\partial t}=-\nabla\cdot[(e+p+\frac{1}{2\mu_0}|\boldsymbol{B}|^{2})\boldsymbol{v}]+\nabla\cdot[\frac{1}{\mu_0}(\boldsymbol{v}\cdot\boldsymbol{B})\boldsymbol{B}]
+\nabla\cdot[\frac{\eta}{\mu_0}\boldsymbol{B}\times(\nabla\times\boldsymbol{B})]-\nabla\cdot\boldsymbol{F_\mathrm{c}}
+\rho\boldsymbol{g}\cdot\boldsymbol{v}+L_{\mathrm{rad}},
\end{equation}
\begin{equation}
\frac{\partial\boldsymbol{B}}{\partial t}=\nabla\times(\boldsymbol{v}\times\boldsymbol{B}-\eta\nabla\times\boldsymbol{B}),
\end{equation}
\begin{equation}
e=\frac{p}{\gamma-1}+\frac{1}{2}\rho|\boldsymbol{v}|^2+\frac{1}{2\mu_0}|\boldsymbol{B}|^{2},
\end{equation}
\begin{equation}
p=\frac{(1.1+Y_{\mathrm{iH}}+0.1Y_{\mathrm{iHe}})\rho}{1.4m_\mathrm{i}}k_\mathrm BT,
\end{equation}
where $\rho$, $\boldsymbol{v}$, $e$, $\boldsymbol{B}$, $m_{\mathrm i}$, $p$, $T$ are mass density, fluid velocity, total energy density, magnetic field, the mass of a proton, thermal pressure, and temperature, respectively. The gravitational acceleration of the sun is $\boldsymbol{g}=273.93\ \mathrm m\ \mathrm s^{-2}$, $\mu_0=4\pi\times10^{-7}\ \mathrm N\  \mathrm A^{-2}$ is the vacuum permeability, $k_\mathrm B=1.3806\times10^{-23}\ \mathrm J\ \mathrm K^{-1}$ is the Boltzmann constant, $I$ is the unit tensor, $\gamma = 5/3 $ is the ratio of specific heats. $L_\mathrm{rad}$ and $\nabla\cdot F_\mathrm c $ refer to the radiative cooling and heat conduction. $Y_{\mathrm{iH}}$ and $Y_{\mathrm{iHe}}$ are the ionization fractions of hydrogen and helium, respectively.

\   The number density of helium is considered to be 10\% of the total hydrogen, and a simple temperature-dependent ionization degree of helium $Y_{\mathrm{iHe}}$  is deduced, according to the experiences by solving the radiative transfer equations \citep{Ni2022}. In the upper convection zone and photosphere, we assume the plasmas are in the local thermodynamic equilibrium state, and the Saha equations that depend on temperature and density are applied for the hydrogen ionization fraction $Y_\mathrm i$. In the chromosphere, $Y_\mathrm i$ is obtained from the temperature-dependent table provided by \citet{CarlssonLeenaarts2012}. In the corona, we assume the plasmas are fully ionized. The ionization fractions in the simulation are updated every half time step according to the local plasma parameters.
The Spitzer type magnetic diffusion coefficient (e.g., \citeauthor{Collados2012} \citeyear{Collados2012}, \citeauthor{Ni2022} \citeyear{Ni2022}) is given by:
\begin{equation}
\eta=\eta_{\mathrm{ei}}+\eta_{\mathrm{en}}=\frac{m_\mathrm e\nu_\mathrm{ei}}{e_{\mathrm c}^2n_{\mathrm e}\mu_0}+\frac{m_{\mathrm e}\nu_{\mathrm{en}}}{e_{\mathrm c}^2n_{\mathrm e}\mu_0},
\end{equation}
where $m_\mathrm e$, $e_\mathrm c$, $n_\mathrm e$ are mass of electron, electron charge and electron density, $\nu_\mathrm {ei}$ and $\nu_\mathrm{en}$ are frequencies of electron-ion and electron-neutral collisions, respectively:
\begin{equation}
\nu_{\mathrm{ei}}=\frac{n_\mathrm ee_{\mathrm c}^4\Lambda}{3m_\mathrm e^2\epsilon_{0}^2}(\frac{m_\mathrm e}{2\pi k_\mathrm BT})^{\frac{3}{2}},
\end{equation}
\begin{equation}
\nu_{\mathrm{en}}=n_\mathrm n\sqrt{\frac{8 k_\mathrm BT}{\pi m_{\mathrm{en}}}}\sigma_{\mathrm{en}},
\end{equation}
where $\epsilon_{0}$ is the permittivity of vacuum, $\Lambda$ is the Coulomb logarithm, $n_\mathrm n$ is the number density of the neutral particles, and $\sigma_{\mathrm{en}}$ is the collision cross section. Since the neutral particle $m_\mathrm n$ is much greater than the electron mass $m_\mathrm e$, we can get   $m_{\mathrm{en}}=m_\mathrm em_\mathrm n/(m_\mathrm e+m_\mathrm n)\cong m_\mathrm e$. The expression of $\Lambda$ is given by:
\begin{equation}
\Lambda=23.4-1.15\mathrm{log}_{10}n_\mathrm e+3.45\mathrm{log}_{10}T,
\end{equation}
with $n_\mathrm e$ expressed in cgs units and $T$ in eV. The collision frequency $v_{\mathrm{en}}$ is contributed by collisions between electrons and neutral hydrogen and collisions between electrons and neutral helium, respectively. Since the smallest grid size is about $5.68$ km, the numerical diffusivity is close to the physical diffusivity in the low solar atmosphere, but it is much larger than the calculated Spitzer diffusivity ($\sim1\ \rm m^2\ \rm s^{-1}$) in the corona. We have measured the numerical diffusivity by using a similar method to that in the previous works (e.g., \citeauthor{Ni2015} \citeyear{Ni2015}, \citeauthor{Shen2011} \citeyear{Shen2011}).

\subsection{Radiation and thermal conduction models} \label{subsec:tables}
\   The energy transport from the interior of the Sun to the outer atmosphere involves the radiative transfer process. Including the radiative transfer process is indispensable for simulating the realistic convection zone and solar atmosphere. However, solving the radiative transfer equations and MHD equations together at each time step costs a significant amount of computational time, especially for the chromosphere that considering the non-local thermodynamic equilibrium state. In this work, we have used different simple radiative cooling models from the upper convection zone to the low corona.

\   In the convection zone, the mean free path of a photon is much smaller than the local pressure scale height, and we treat the radiative transport process in this region with the diffusion approximation, the formula is as follows:
\begin{equation}
L_{\mathrm{rad1}}=\nabla\cdot(k_{\mathrm{rad}}\nabla T),
\end{equation}
where $k_{\mathrm{rad}}=16\sigma T^3/(3k_\mathrm R)$ is the coefficient of conduction for radiative transport \citep{Kippenhahn1994}, $\sigma=5.67\times 10^{-8}\ \mathrm J\ \mathrm s^{-1}\ \mathrm m^{-2}\ \mathrm K^{-4}$ is the Stefan-Boltzmann constant, and $k_\mathrm R$ is the atomic Rosseland mean opacity of the sun-like star model from the OPAL tables \citep{Rogers1992}.
\   In the photosphere, an approximate model proposed by \citet{Abbett2012} is applied:
 \begin{equation}
L_{\mathrm{rad2}}=-2k^\mathrm B\rho\sigma T^4E_2(\tau^\mathrm B),
\end{equation}
where $k^\mathrm B$ is the Planck-averaged opacity, $\tau^\mathrm B$ is the optical depth calculated from this opacity, and $E_2$ is a function depending on the optical depth  $\tau^\mathrm B$. We also use the Rosseland mean opacity to get the value of $k^\mathrm B$ and then the corresponding value of $\tau^\mathrm B$ is obtained by integration along the vertical direction.
We use the widely accepted \citet{CarlssonLeenaarts2012} model for the chromosphere. This model focuses on the effect of radiative cooling in several important chromospheric spectral lines, such as H I, Ca II, and Mg II. The data from RADYN and Bifrost are used to fit the table of associated parameters \citep{Carlsson2002, Gudiksen2011}. The formula of this model is given by:
 \begin{equation}
L_{\mathrm{rad3}}=-\sum_{\mathrm{X=H,Mg,Ca}}L_{\mathrm X_m}(T)E_{\mathrm X_m}(\tau)\frac{N_{\mathrm X_m}}{N_{\mathrm X}}(T)A_\mathrm X\frac{N_\mathrm H}{\rho}n_\mathrm e\rho,
\end{equation}
where $L_{\mathrm X_m}$ is the optically thin radiative loss function per electron and per particle of element X in ionization stage $m$, $E_{\mathrm X_m}(\tau)$ is the photon escape probability as function of the depth parameter $\tau$ that depends on the column mass, $\frac{N_{\mathrm X_m}}{N_\mathrm X}$  is the fraction of element X in ionization stage $m$, and $A_\mathrm X$ corresponds to the abundance of element X. In this simulation, X comprises H, Ca, and Mg.  Most of the parameters can be determined by consulting the table offered by \citet{CarlssonLeenaarts2012}. According to the solar atmospheric model \citep{Avrett2008}, $A_\mathrm H=1$, $A_{\mathrm{Mg}}=3.885\times10^{-5}$, $A_{\mathrm{Ca}}=2.042\times10^{-6}$.

\   The radiative cooling model for the corona is also given by \citet{CarlssonLeenaarts2012} as:
 \begin{equation}
L_{\mathrm{rad4}}=-n_\mathrm Hn_\mathrm ef(T),
\end{equation}
where $n_\mathrm H$ is the number density of hydrogen, $f(T)$ is a function of temperature that is derived based on the calculation results from the Bifrost code \citep{Carlsson2002, Gudiksen2011}.
The models applied in different layers are separated by the value of the neutral hydrogen column mass. Therefore, the radiative cooling formula in the whole computation is as follows:
 \begin{equation}
L_{\mathrm{rad}}=\begin{cases}L_{\mathrm{rad1}},\ m_\mathrm c >m_{\mathrm{c1}}\\
L_{\mathrm{rad2}},\ m_{\mathrm{c2}}<m_\mathrm c\leq m_{\mathrm{c1}}\\
L_{\mathrm{rad3}},\ m_{\mathrm{c3}}<m_\mathrm c\leq m_{\mathrm{c2}}\\
L_{\mathrm{rad4}},\ m_\mathrm c\leq m_{\mathrm{c3}},
\end{cases}
\end{equation}
where $m_{\mathrm{c1}}=10^{1.56}\ \mathrm {kg\ m}^{-2}$, $m_{\mathrm{c2}}=10^{-1.6}\ \mathrm {kg\ m}^{-2}$, $m_{\mathrm{c3}}=10^{-4.7}\ \mathrm {kg\ m}^{-2}$, $m_\mathrm c$ is the column mass calculated by:
\begin{equation}
m_{\mathrm{c}}(x,y)=\int _y^\infty \rho(x,y)dy,
\end{equation}
here $\rho(x,y)$ is the mass density at grid $(x, y)$. The radiative cooling models for the chromosphere and corona applied in this work have been widely used in the previous radiative MHD simulations (e.g., \citeauthor{Martinez-Sykora2017} \citeyear{Martinez-Sykora2017}, \citeauthor{Hansteen2017} \citeyear{Hansteen2017}). The approximate treatment of the optically thick radiative cooling model applied in the photosphere in this work has been tested \citep{Abbett2012}, and it can successfully reproduce a stable convection pattern, which is close to the results by solving the radiative transfer equations. In the convection zone with high opacity, the model $L_{\mathrm{rad1}}$ can also approximately present the radiative environment there. These applied simple radiative cooling models greatly save the computational costs, and the dynamics in the computational domain are still well resolved. However, solving the radiative transfer equations at each time step can make the synthesized spectra and spectral line profiles for small-scale events in the partially ionized low solar atmosphere closer to the observational results (e.g., \citeauthor{Hansteen2019} \citeyear{Hansteen2019}, \citeauthor{Martinez-Sykora2017} \citeyear{Martinez-Sykora2017}).

\   We include the heat conduction term along the magnetic field lines in this work. The heat flux $F_\mathrm c$ is given by:
\begin{equation}
F_{\mathrm c}=-k_{\parallel}(\nabla T\cdot\boldsymbol{\hat {B}})\boldsymbol{\hat {B}},
\end{equation}
where  $\boldsymbol{\hat {B}}= \boldsymbol B/|\boldsymbol B|$ is the unit vector in the direction of magnetic field, $k_{\parallel}$ is the parallel conductivity coefficient which is given by Spitzer theory \citep{Spitzer1962}:
\begin{equation}
k_{\parallel_{\mathrm{sp}}}=\frac{1.84\times10^{-10}}{\Lambda}T^{\frac{5}{2}}.
\end{equation}

\subsection{Initial setups and boundary conditions} \label{subsec:tables}
\begin{figure}[!h]
\centering
\includegraphics[width=8.6cm]{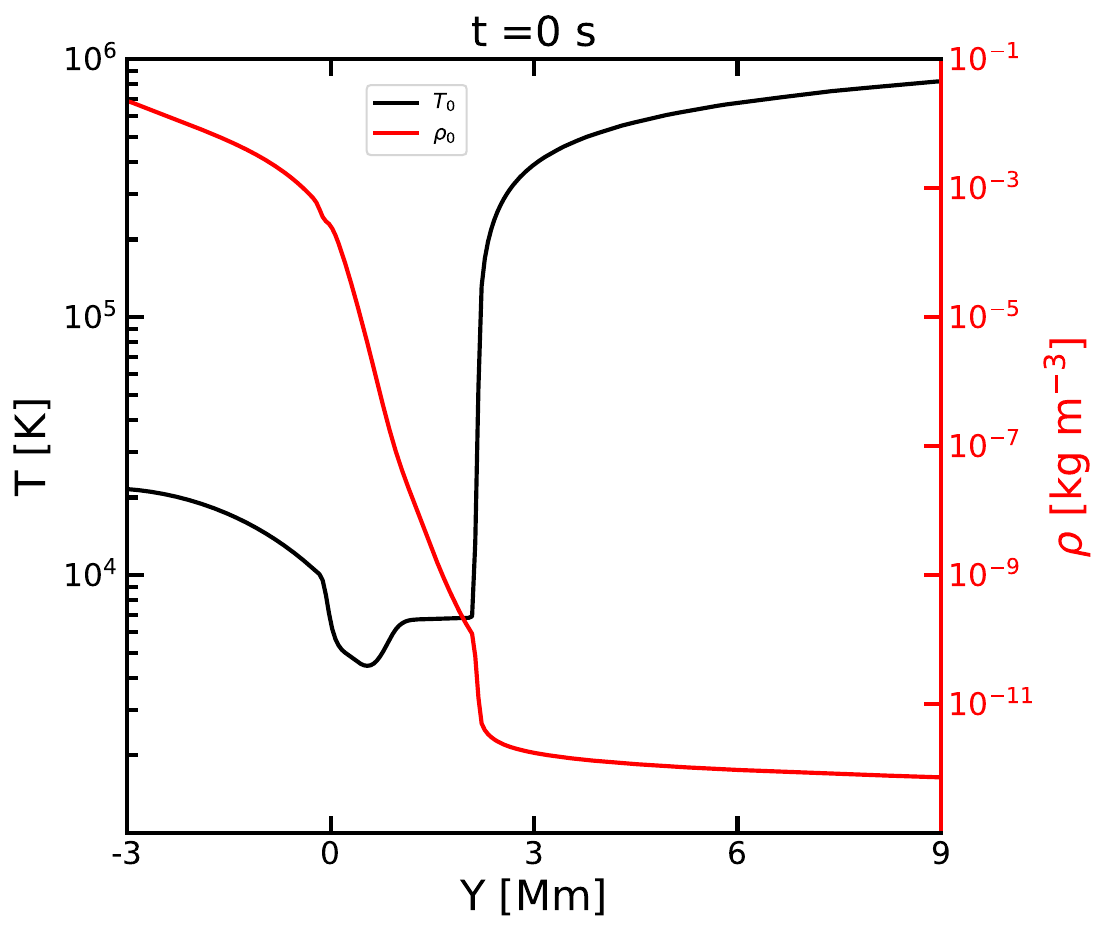}
\caption{Initial temperature (black solid) and density (red solid) profiles in height. \label{fig1}}
\end{figure}
\   The simulation domain covers the region within $-10<x<10$\ Mm and $-3<y<9$\ Mm, where x is in the horizontal direction, y is in the direction vertical to the solar surface. An initial grid of dimension $256 \times 512$ spans the simulation domain, and a 2-level adaptive mesh refinement is applied, with each level of mesh refinement doubling the number of grid points. Therefore, the smallest grid size in the x and y directions is 19.53 km and 5.86 km, respectively. We use the periodic boundary conditions at the left and right boundaries, and the inflow and outflow boundary conditions are applied at the bottom and the top boundaries, respectively. At both the bottom and top boundaries, the gradients of the thermal energy, mass density, and parallel components of the magnetic field are set to zero, and the perpendicular component of the magnetic field is obtained by divergence-free extrapolation.
The gradients of velocities in the x and z-direction also vanish at the top and bottom boundaries by assuming $\partial v_x/\partial y = 0$, $\partial v_z/\partial y = 0$. At the bottom boundary, the fluid is only allowed to flow into the computation domain by assuming:
\begin{equation}
v_{ybg}=\begin{cases}v_{ybl},\ if\ v_{ybl}>0\\
-v_{ybl},\ if\ v_{ybl}<0,
\end{cases}
\end{equation}
where $v_{ybg}$ and $v_{ybl}$ separately represent the velocities at the two bottom ghost layers and at the first two layers inside the simulation domain in the y-direction. At the up boundary, the fluid is only allowed to flow out the computation domain by assuming:
\begin{equation}
v_{yug}=\begin{cases}v_{yul},\ if\ v_{yul}>0\\
-v_{yul},\ if\ v_{yul}<0,
\end{cases}
\end{equation}
where $v_{yug}$ and $v_{yul}$ separately represent the velocities at the two up ghost layers and the last two layers inside the simulation domain in the y-direction. We noted that the reconnection current sheet in the simulation domain is far from the left and right boundaries, which prevents the effects of boundary conditions on the reconnection process.

\   The initial uniform magnetic field is set as $B_y=10$ G. The standard solar model and C7 solar atmospheric model \citep{Avrett2008} is applied to set the plasma parameters in the convection zone and in the solar atmosphere, the specific density and temperature distributions along the y-direction are presented in Fig. \ref{fig1}. We assume that all variables are uniformly distributed in the x-direction at the initial moment, except that we have initiated a small perturbation in density at around $y = -2.6$ Mm in the convection zone to trigger the faster evolution of the whole system, which is also widely applied in most existing radiative MHD simulations. Since the initial state does not satisfy the hydrostatic balance, the system underwent self-adjustment in the early stage and eventually reached a dynamic equilibrium state after 2000 s. The convective and turbulent motions are self-consistently generated. In many recent radiative MHD simulations (e.g., \citeauthor{Iijima2015} \citeyear{Iijima2015}, \citeauthor{Martinez2017} \citeyear{Martinez2017}), such initial setups are also applied, and the simulated system went through a similar evolution process. These initial setups are different from those in the previous works of numerical studies of oscillation reconnection (e.g., \citeauthor{McLaughlin2009} \citeyear{McLaughlin2009}, \citeauthor{Murray2009} \citeyear{Murray2009}).
At $t=2048.23$ s, we insert a magnetic flux rope in the convection zone, as shown in Fig. \ref{fig2}(a). The convective and turbulent motions cause this magnetic flux rope emerging into the upper solar atmosphere. The inserted rope is the Gold-Hoyle like flux tube model \citep{Magara2001}, and its magnetic field distributions are as follows:
\begin{equation}
B_x=-B_1\frac{b(y-y_0)}{10^6}exp(-\frac{(x-x_0)^2+(y-y_0)^2}{10^{12}c}),
\end{equation}
\begin{equation}
B_y=B_1\frac{b(x-x_0)}{10^6}exp(-\frac{(x-x_0)^2+(y-y_0)^2}{10^{12}c}),
\end{equation}
\begin{equation}
B_z=B_1exp(-\frac{(x-x_0)^2+(y-y_0)^2}{10^{12}c}),
\end{equation}
where $x_0$ and $y_0$ are the coordinates of the center of the magnetic flux rope, $B_1$ is the magnetic field strength at the center of the magnetic flux rope, $b$ is the twist number, and $c$ is the decay factor. Here, we set $x_0=2\times10^6 \ \mathrm m$, $y_0=-10^6 \ \mathrm m$, $B_1=6400 \ \mathrm G$, $b=3.0$, $c=0.055$.
\section{Results}
\subsection{The oscillatory magnetic reconnection with the reversal of the current sheet orientation}
\begin{figure}[!h]
\centering
\includegraphics[width=18cm]{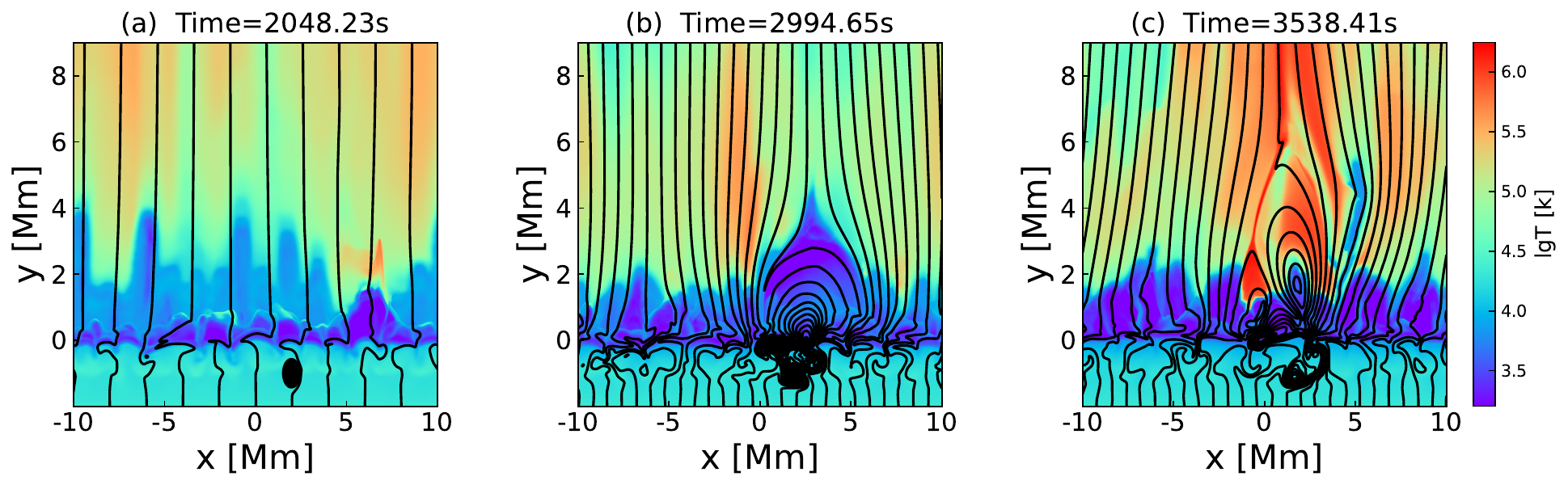}
\caption{The distribution of temperature, overlaid with magnetic fieldlines at $t=2048$, 2994, and 3538 s are presented. \label{fig2}}
\end{figure}

\begin{figure}[!h]
\centering
\includegraphics[width=18cm]{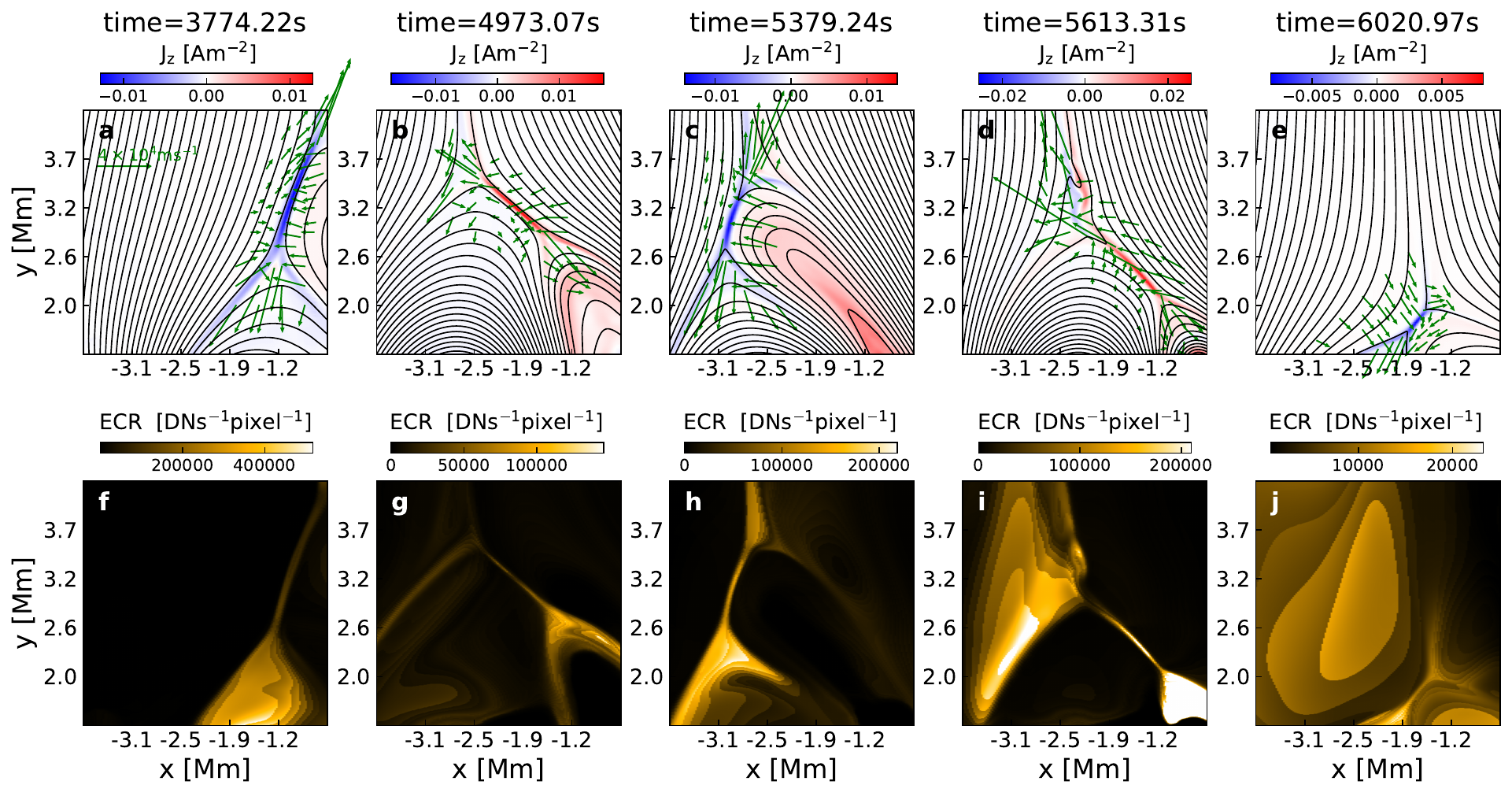}
\caption{(a)-(e) displays the current density distributions in the z direction at five different times, with the black solid line representing the magnetic field lines, the green arrows represent the velocity near the reconnection region; (f)-(j) show the corresponding synthesized images in the AIA 17.1 nm pass band. \label{fig3}}
\end{figure}
\   The magnetic flux rope in the convection zone rises to the base of the photosphere due to the magnetic buoyancy and convective processes. Subsequently, it emerges into the corona by triggering further instabilities \citep{Cheung2014}. In the gravitationally stratified solar atmosphere, the steep gradient of the pressure scale height causes the emerging magnetic flux rope to expand both horizontally and vertically. Such an expansion process compresses the ambient magnetic field, as shown in Fig. \ref{fig2}(b). Since the magnetic field on the left side of the emerging rope is antiparallel to the background field, a current sheet forms at the interaction region, and the magnetic reconnection occurs within the current sheet. Figure. \ref{fig2}(c) shows that the temperature of the current sheet reaches a million Kelvin. The post-reconnection loops are continually generated in the downward reconnection outflow region.

\   During the reconnection process, the length of the current sheet gradually decreases to zero and then elongates in its orthogonal direction. Such transitions of the current sheet orientation occur repeatedly in our simulation, accompanied by reversals between the reconnection inflow and outflow regions. For example, the current sheet extends from the lower-left to the upper-right corner at $t=3774$ s (see Fig. \ref{fig3}a), then its orientation transits and it extends from the lower-right to the upper-left corner at $t=4973$ s (see Fig. \ref{fig3}b).  According to the definitions by \citet{Murray2009} and \citet{McLaughlin2012b}, the duration of a single current sheet orientation is considered as an individual reconnection phase, while the interval between two adjacent phases constitutes a full oscillation period. Figure \ref{fig3}, a-e, illustrates the evolution of the current $J_z$ during the reconnection phases from $t=3774$ s to $t=6020$ s. Reversals of the reconnection inflow and outflow regions lead to polarity reversals of the current $J_z$ in the reconnection region. Based on the Chianti package \citep{DelZanna2015}, we have synthesized the images in the AIA 17.1 nm band (e.g., \citeauthor{Ni2017} \citeyear{Ni2017}) by using the calculated density and temperature distributions from our simulation results. The synthesized images show bright bifurcated structures (see Fig. \ref{fig3}, f-j), which are consistent with the corresponding current sheet structures from observations \citep{Hong2019, Xue2019, Sun2023}.

\   The entire oscillatory reconnection process spans from $t=3357$ s to $t=9130$ s, featuring 41 reconnection reversals in our simulations. The duration of individual reconnection phases is 1502, 299, 314, 353, 31, 120, 75, 91, 100, 70, 41, 154, 19, 79, 108, 55, 67, 13, 33, 62, 57, 64, 238, 96, 243, 74, 11, 15, 207, 103, 13, 17, 330, 63, 108, 34, 105, 78, 41, 118, and 45 seconds, corresponding to 40 oscillation periods. Therefore, the first oscillation period is about 30 minutes, which is the same as the period of the observed oscillatory magnetic reconnection with the reversal of current sheet orientation by \citet{Hong2019}, but it is shorter than the observed one by \citet{Xue2019}. Such a long period of 30 minutes has not been displayed in the previous papers, which only show periods of tens of seconds to a few minutes (e.g., \citeauthor{Murray2009} \citeyear{Murray2009}, \citeauthor{McLaughlin2012a} \citeyear{McLaughlin2012a}, \citeauthor{Karampelas2022a} \citeyear{Karampelas2022a}, \citeauthor{Karampelas2022b} \citeyear{Karampelas2022b}). The periods of the reversal of the current sheet orientation can be significantly affected by the plasma environments, the strength and configurations of background and emerged magnetic fields \citep{Karampelas2022b, Karampelas2023, Schiavo2024a, Schiavo2024b, Talbot2024}. We should also note that the reversal of current sheet orientation with a shorter period of several minutes or less might also occur in the solar atmosphere, but the limited resolution of the existing solar telescopes causes difficulties in observing such short-period phenomena.
\subsection{The triggering mechanism of oscillatory reconnection}
\begin{figure}[!h]
\centering
\includegraphics[width=18cm]{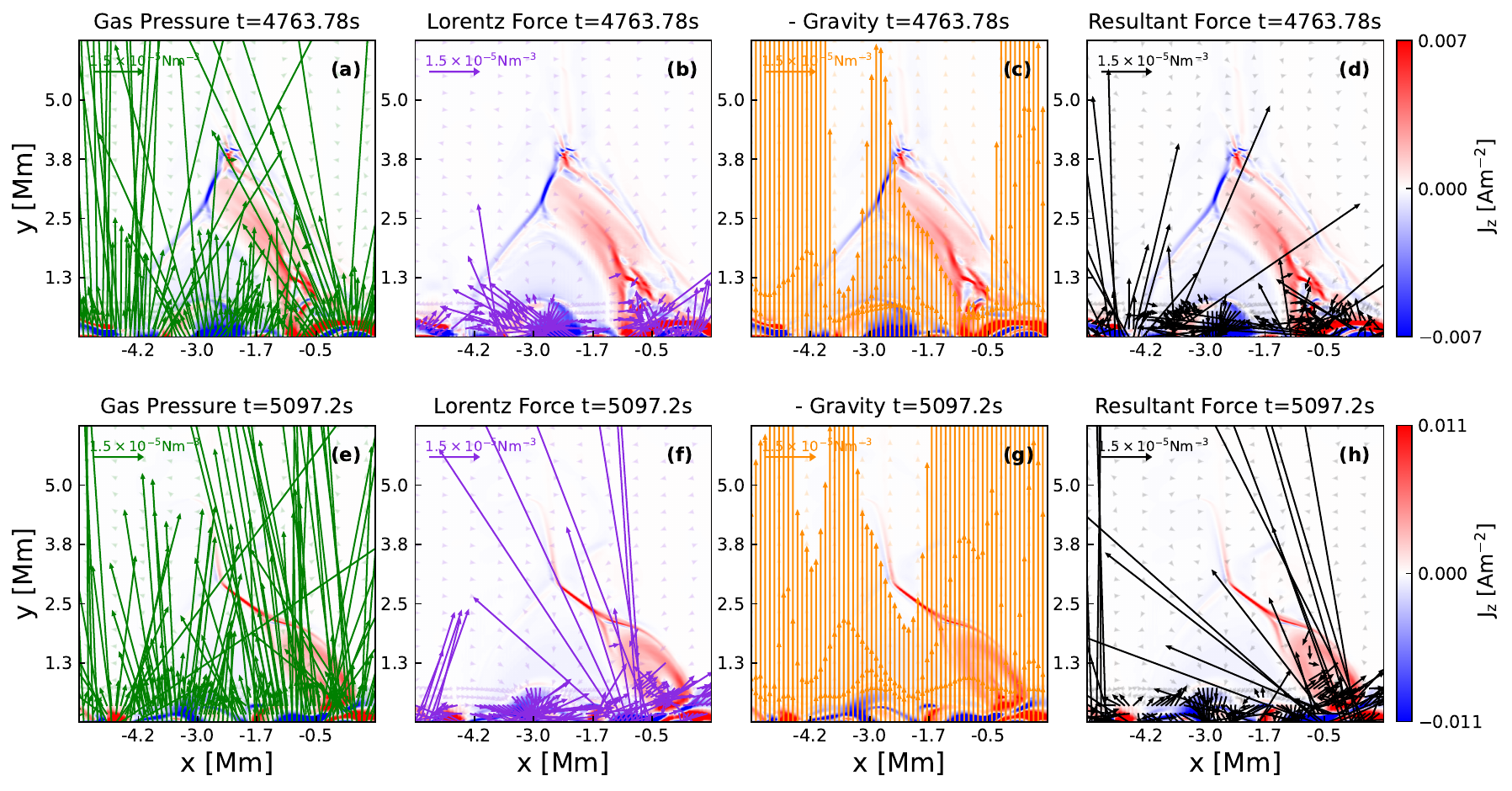}
\caption{The arrows with different colors in (a)-(d) represent the plasma pressure gradient, the Lorentz force per unit volume, and the negative gravity per unit volume better shows the effect of gravity originating from the photosphere and the resultant force per unit volume at $t=4763.78$ s. The arrows in (e)-(h) are the same but at $t=5097.2$ s. The values less than $1.5\times 10^{-6}\  \mathrm N\ \mathrm m^{-3}$ are represented in a faded manner. The blue-red colors in these panels represent the current density $J_z$. \label{fig4}}
\end{figure}

\   To determine the mechanism causing the current sheet reversals, we have analyzed the time evolution of different force distributions, including the gas pressure gradient, the Lorentz force, and gravity in the simulation domain. The arrows with different colors in Fig. \ref{fig4} represent the plasma pressure gradient, the Lorentz force per unit volume, the negative gravity per unit volume, and the resultant force per unit volume at two different times, respectively. The values less than $1.5\times 10^{-6}\ \mathrm N\ \mathrm m^{-3}$ are represented in a faded manner. The blue short current sheet shown in Fig. \ref{fig4}(a)-(d) is at the moment when it is about to disappear, and the new red current sheet with a reversed orientation is about to appear. Though the continuous reconnection leads to the accumulation of plasma and magnetic flux in the outflow region, we find that the resulting force in the outflow region of this blue current sheet is very weak, which is unlikely to be the main reason for driving the oscillatory magnetic reconnection. Near the surface of the Sun, all forces increase dramatically, and they mainly point toward the upper right direction, which then squeezes this current sheet and drives the formation of a red current sheet with a reversed orientation in the next stage.

\   During the second phase of the oscillatory magnetic reconnection, a red current sheet extending orthogonally to the previous blue one appears, and the reconnection outflow/inflow regions are reversed. The competition among the Lorentz force, pressure, and gravity caused by the emerging magnetic fields and plasma on both sides also begins to change. Figure. \ref{fig4}(e)-(h) display the distribution of different forces at the moment when the second reconnection phase is about to end and the third reconnection phase is about to begin. The resulting force by the reconnection outflows is still not significant during this phase. The emerging plasma from the convection zone results in significant upward gas pressure and downward gravity near the solar surface, while the upward Lorentz force on the left side of the bottom region is much weaker. However, the emergence and expansion of the inserted magnetic flux rope led to a significant generation of Lorentz force and gas pressure in the lower right region. Finally, there is a significant net force along the upper left direction in this region below the red current sheet. This force compresses the red current sheet and causes the lower right outflow region to approach the upper left outflow region, and then the red current sheet gradually shortens to zero.

\    From the above analyses, we can find that the main factor leading to the reversal of the current sheet's orientation is the variation in plasma pressure gradient, Lorentz force, and gravity beneath the lower outflow region of reconnection. These variations result from the emergence of plasma and magnetic fields from the convection zone into the solar atmosphere. Therefore, the driving mechanism of the oscillatory reconnection in our simulation is the quasi-periodic external force provided by the emergence of plasma and magnetic fields from the convection zone. This driving mechanism is different from the previous simulation works, which reported that the driving mechanism of oscillatory magnetic reconnection originated from magnetic reconnection itself. (e.g., \citeauthor{McLaughlin2009} \citeyear{McLaughlin2009}, \citeauthor{Murray2009} \citeyear{Murray2009}, \citeauthor{Thurgood2017} \citeyear{Thurgood2017}).
\subsection{The intermittent jets caused by the oscillatory magnetic reconnection}
\begin{figure}[!h]
\centering
\includegraphics[width=18cm]{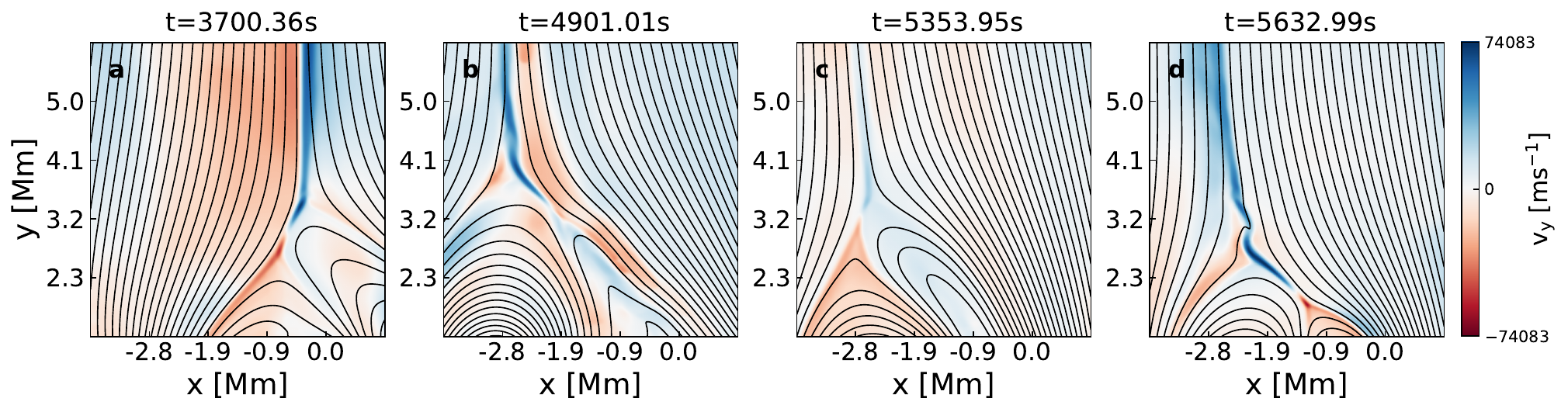}
\caption{(a)-(d) show the distributions of the vertical velocity $v_\mathrm y$ at four different reconnection stages. \label{fig5}}
\end{figure}

\   The alternating changes in the inflow and outflow regions during the oscillatory reconnection process direct the upward outflows to various locations with open field lines. As shown in Fig. \ref{fig5}a, magnetic reconnection leads to two opposing high-speed outflows along the current sheet. The downward outflow in the lower part of the current sheet enters the post-reconnected loop region at the bottom, further forming two downward-diverting flows along both sides of the loop. The upper reconnection outflow is ejected toward the upper right region with open field lines, where the flow is further divided into upward and downward flows. The upward component creates a collimated jet along the vertical open field lines at around $x = 0$ Mm. Due to the change of the reconnection outflow region during the second reconnection phase, the collimated jet along the y-direction caused by the upward outflow moves to the location near $x = -2.8$ Mm at $t=5632.99$ s.

\   Similar changes occur in the subsequent reconnection phases. The reversal of the current sheet orientation alters the positions of the reconnection outflows, and then the upward outflows enter different regions with open field lines, and a series of parallel jets is created. These jets are all ejected along the y-direction, but they originate from different locations in the x-direction due to the changes of reconnection outflow regions; they can well explain the observed parallel shifting jets in different wavelengths (e.g., \citeauthor{Joshi2017}\citeyear{Joshi2017}, \citeauthor{Mulay2017} \citeyear{Mulay2017}). If the background magnetic fields are more complex than simple y-directional magnetic fields, the reconnection outflows caused by the oscillation reconnection process might enter into different zones with magnetic fields oriented in various directions. Then, the intermittent jets along different directions will be formed \citep{Cai2024}.

\   Since the length of the current sheet in an oscillatory reconnection process is constantly growing or shrinking in the line of sight direction, the outflow positions will continuously change. If the observational instruments have sufficient high resolutions, the horizontal displacements of the intermittent jets will be observed.

\subsection{The oscillations of reconnection rates}
\begin{figure}[!h]
\plotone{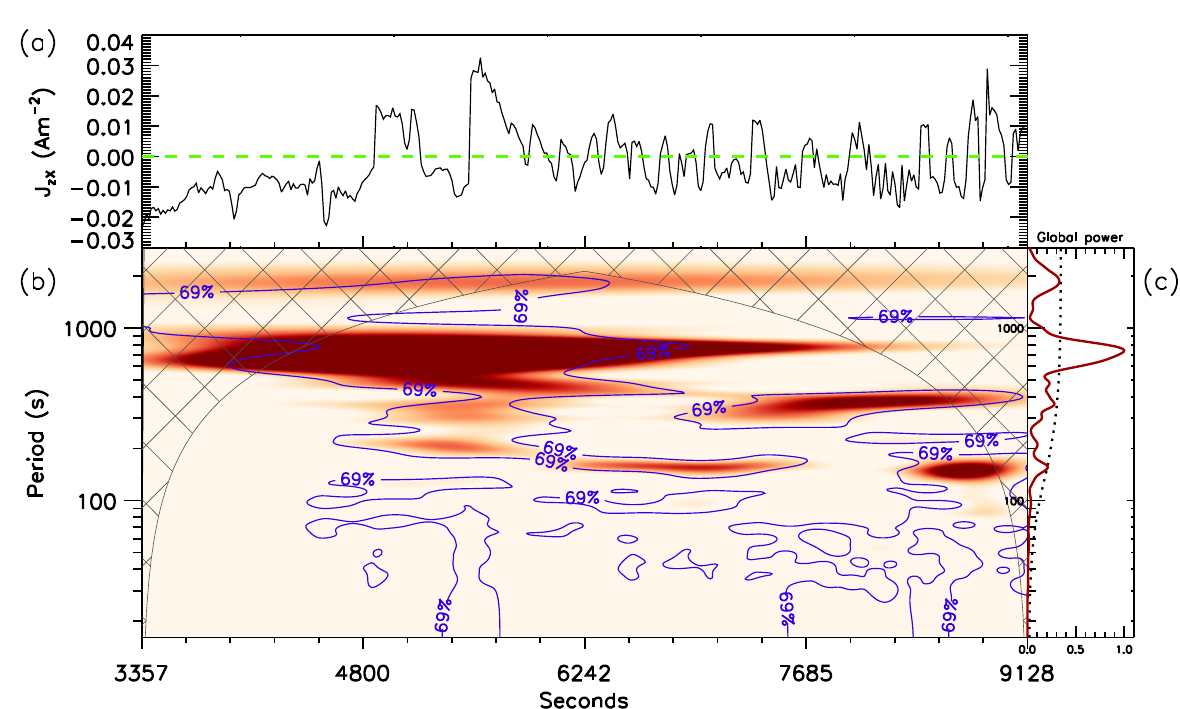}
\caption{(a) The time evolutions of the current density at the main reconnetcion X-point ($J_{z\mathrm X}$). (b) The wavelet profile for $J_{z\mathrm X}$. The horizontal green dashed line represents $J_z=0$. The regions within the blue contours have a confidence level $\geq 69$\%. (c) The solid line is the global wavelet power for $J_{z\mathrm X}$. The dashed line is the 69\% confidence level for the global wavelet power. \label{fig6}}
\end{figure}
\   We have measured the time-dependent current density $J_{z\mathrm X}$ at the main reconnection X-point, which also shows the obvious oscillation behaviors. One should note that $\eta J_{z\mathrm X}$  represents the value of the reconnection rate at the main X-point \citep{Leake2013, Ni2018, Liu2023}. As shown in Figure \ref{fig6}(a), the multiple peaks of $J_{z\mathrm X}$  usually occur during each reconnection phase before the current sheet orientation reverses for the next phase. Figure \ref{fig6}(b) shows the corresponding wavelet spectrum obtained by using the method described by \citet{Torrence1998}. The regions within the blue contours have a confidence level $\geq$ 69\%. Several main oscillation periods are distributed at around 1800 s, 500 s - 900 s, 160 s, and 300 s. These periods correspond to the first oscillation period of 1801 s during the oscillatory reconnection process with the reversal of the current sheet orientation, as well as the shorter periods for subsequent multiple current sheet orientation reversals.

\   In addition, the short periods distributed between 100 s and 400 s also cover the oscillation period of the reconnection rate at the main X-point, which does not necessarily require the reversal of the current sheet orientation.  Such a periodicity in the reconnection rate is close to the periodicity of the 2 - 5 minute quasi-periodic inflow from the convection zone observed in radiative MHD simulations \citep{Kitiashvili2013} and the 2.5 - 5 minute oscillation periods of flare emission intensity in different wavelengths (e.g., \citeauthor{Innes1997} \citeyear{Innes1997}, \citeauthor{Peter2014} \citeyear{Peter2014}, \citeauthor{Tian2016} \citeyear{Tian2016}, \citeauthor{Young2018} \citeyear{Young2018}, \citeauthor{Ning2014} \citeyear{Ning2014}, \citeauthor{Li2020} \citeyear{Li2020}, \citeauthor{Hayes2020} \citeyear{Hayes2020}, \citeauthor{Zimovets2021} \citeyear{Zimovets2021}). Such a periodicity may also relate to the widely observed 3-minute and 5-minute p-mode oscillation events \citep{Chen2006}, which result from the continuous convective motions and the MHD waves in the convection zone \citep{Bahng1963}. Helioseismic investigations have shown a strong correlation between photospheric velocity oscillation frequencies and granulation frequencies \citep{Frazier1968}. We should also note that there are some periods shorter than 20 s during the reconnection process, since the time interval applied in this wavelet analysis program is approximately 15 s, these short periods are not shown in Fig. \ref{fig6}(b). As described in the last subsection, the shortest period is only about 26 s for the reversal of the current sheet orientation. The previous works showed that turbulent reconnection mediated with plasmoids can cause the oscillation period of the reconnection rate to be smaller than 1 minute \citep{Ni2015, Ye2020}.

\section{Conclutions and Discussion}
\   We have performed 2.5D Radiative MHD simulations to study the physical process in which a magnetic flux rope emerges from the convection zone into the solar atmosphere and reconnects with the background magnetic field. Based on the previous work of \citet{Cheng2024}, the radiative cooling models for the convection zone and corona are further incorporated. The convective motions in the convection zone then cause the inserted magnetic flux rope there to emerge self-consistently into the solar atmosphere. Since the direction of the emerging magnetic loop on the left side is opposite to that of the vertical background magnetic field, a current sheet forms at their interface. As magnetic reconnection proceeds, the length of the current sheet along the lower-left to upper-right direction gradually shortens until it nearly disappears, and a new current sheet extends in the direction perpendicular to the previous one, the inflow and outflow regions are also reversed. Such a phenomenon is known as oscillatory magnetic reconnection. We have synthesized the images of the current sheet in the AIA 17.1 nm band (e.g., \citeauthor{Ni2017} \citeyear{Ni2017}) by using the calculated density and temperature distributions from our simulations. The radiation features closely resemble the observed current sheet structures \citep{Hong2019, Xue2019, Sun2023}. The main conclusions are summarized below:

\   1. The oscillatory magnetic reconnection process includes 40 oscillation periods, and the first period is the longest one and it is about 30 minutes, which is consistent with the previous observational results of oscillatory magnetic reconnection with the reversal of the current sheet orientation \citep{Hong2019}. The shortest period we find is about 26 s.

\   2. When plasma and magnetic fields emerge from the convection zone to the solar atmosphere, they alter the plasma pressure gradient, Lorentz force, and gravity below the lower outflow region of reconnection. The quasi-periodic changes of external forces resulting from convective and turbulent motions are the primary cause of the current sheet's orientation reversal. Such a driving mechanism of oscillatory magnetic reconnection is different from all the previous simulation works (e.g., \citeauthor{McLaughlin2009} \citeyear{McLaughlin2009}, \citeauthor{Murray2009} \citeyear{Murray2009}, \citeauthor{Thurgood2017} \citeyear{Thurgood2017}).

\   3. The alternating changes of reconnection inflow and outflow regions cause the reconnection outflows to enter into different regions with open magnetic field lines. The shifting of the upward outflows along the horizontal direction provides a good explanation for the observed parallel shifting jets at different locations near the same reconnection region (e.g., \citeauthor{Joshi2017} \citeyear{Joshi2017}, \citeauthor{Mulay2017} \citeyear{Mulay2017}).

\   4. The reconnection rate at the main X-point also oscillates with a period between 100-400 s, and it can well explain the 2.5-5 minute oscillation periods of flare emission intensity in different wavelengths (e.g., \citeauthor{Innes1997} \citeyear{Innes1997}, \citeauthor{Peter2014} \citeyear{Peter2014}, \citeauthor{Tian2016} \citeyear{Tian2016}, \citeauthor{Young2018} \citeyear{Young2018}, \citeauthor{Ning2014} \citeyear{Ning2014}). The oscillation in the reconnection rate usually occurs several times during each reconnection phase before the current sheet orientation reverses for the next phase. Such oscillations may relate to the widely observed 3-minute and 5-minute p-mode oscillation events \citep{Chen2006}, which result from the continuous convective motions and the MHD waves in the convection zone.

\   The artificially imposed pulses are usually applied to trigger the oscillatory magnetic reconnection in the previous MHD simulations (e.g., \citeauthor{McLaughlin2012a} \citeyear{McLaughlin2012a}, \citeauthor{Thurgood2017} \citeyear{Thurgood2017}). Oscillatory magnetic reconnection between the emerged magnetic flux rope and background magnetic fields was reported by \citet{Murray2009}, an initial density perturbation was included to trigger the emergence of the inserted magnetic flux rope below the solar surface. In all the previous works, the continuous reconnection outflows cause the accretion of hot and high density plasmas in the outflow region, where the high plasma pressure results in the altering of the force balance between the outflow and inflow regions and drives the reversal of the current sheet orientation (e.g., \citeauthor{Murray2009} \citeyear{Murray2009}).

\   In this work, the convective and turbulent motions below the solar surface are generated to trigger the emergence of magnetic flux rope, and magnetic reconnection between the emerged and background fields almost occurs in the low corona. The convective and turbulent motions play the main role in causing the subsequent oscillatory magnetic reconnection with the reversal of the current sheet orientation.  Our simulation results also indicate that the emission intensities (corresponding to the reconnection rates) of small brightenings and solar ares \citep{Innes1997, Peter2014, Tian2016, Young2018, Ning2014, Li2020, Hayes2020, Zimovets2021} can be strongly modulated by the convective and turbulent motions in the convective zone. However, we should mention that the modulation effects of convective motions will become weaker if the reconnection events are located at a higher altitude.

\   The longest period of the current sheet orientation reversal is about 30 minutes, that is close to the previous observed period of oscillatory magnetic reconnection by \citet{Hong2019}, but it is shorter than the observed one by \citet{Xue2019}. We should also point out that such a long period has not been shown in all the previous papers (e.g., \citeauthor{Murray2009} \citeyear{Murray2009}, \citeauthor{McLaughlin2012a} \citeyear{McLaughlin2012a}, \citeauthor{Karampelas2022a} \citeyear{Karampelas2022a}, \citeauthor{Karampelas2022b} \citeyear{Karampelas2022b}). However, the reversal of current sheet orientation with a shorter period of several minutes or less might also occur in the solar atmosphere, which needs the future solar telescopes with higher time and space resolutions to prove.


\begin{acknowledgments}
\    This research is supported by the Basic Research of Yunnan Province in China with Grant 202401AS070044; the NSFC Grants 12373060; the Strategic Priority Research Program of the Chinese Academy of Sciences with Grant No. XDB0560000; the National Key R\&D Program of China No.2022YFF0503003 (2022YFF0503000); the National Key R\&D Program of China No. 2022YFF0503804 (2022YFF0503800); the outstanding member of the Youth Innovation Promotion Association CAS (No. Y2021024 ); the International Space Science Institute (ISSI) in Bern, through ISSI International Team project \# 23-586 (Novel Insights Into Bursts, Bombs, and Brightenings in the Solar Atmosphere from Solar Orbiter);  the Yunnan Key Laboratory of Solar Physics and Space Science under the number 202205AG070009. The simulation work was carried out at National Supercomputer Center in Tianjin, and the calculations were performed on Tianhe new generation supercomputer. The numerical data analysis have been done on the Computational Solar Physics Laboratory of Yunnan Observatories.
\end{acknowledgments}


\bibliographystyle{aasjournal}
\bibliography{paperOS}



\end{document}